\documentclass[preprinty]{aastex}
\usepackage{longtable}
\usepackage{graphicx,amssymb,mathrsfs,amsmath}
\usepackage{lscape}
\usepackage{color}

\begin{document}

\title{Gradual Magnetic Evolution of Sunspot Structure and Filament-Corona Dynamics
Associated with the X1.8 Flare in AR 11283}

\author{GUIPING RUAN\altaffilmark{1},
        YAO CHEN\altaffilmark{1},
        and HAIMIN WANG\altaffilmark{2,3}
} \affil{1 Shandong Provincial Key Laboratory of Optical Astronomy
and Solar-Terrestrial Environment, and Institute of Space
Sciences, Shandong University, Weihai 264209, China;
yaochen@sdu.edu.cn} \affil{2 Space Weather Research Laboratory,
Center for Solar-Terrestrial Research, NJIT, Newark, NJ07102, USA}
\affil{3 Big Bear Solar Observatory, 40386 North Shore, Big Bear
City,  CA 92314, USA}

\begin{abstract}

In this paper, we present a study on persistent and gradual
penumbral decay and correlated decline of the  photospheric
transverse field component during 10-20 hours before a major flare
(X1.8) eruption on 2011 September 7. This long-term pre-eruption
behavior is corroborated with the well-imaged pre-flare filament
rising, the consistent expansion of coronal arcades overlying the
filament, as well as the NLFFF modelling results in the
literature. We suggest that both the long-term pre-flare penumbral
decay and the transverse field decline are the photospheric
manifestation of the gradual rise of the coronal filament-flux
rope system. We also suggest that a C3 flare and subsequent
reconnection process preceding the X1.8 flare play an important
role in triggering the later major eruption.

\end{abstract}


\section{Introduction}
Photospheric and corona magnetic field evolution is essential to
the energy build-up and release process of solar eruptions. The
field evolution is closely related to changes of sunspot
structures. It is thus of vital importance to establish the physical
connection between the photospheric magnetic field, the sunspot
evolution, and the coronal dynamics.

Many previous studies have been focusing on flare-induced changes
of the photospheric magnetic field and sunspot structures (e.g.,
Severny 1964; Wang et al., 1994; Spirock et al., 2002; Yurchyshyn
et al., 2004; Wang \& Liu 2010; Wang et al., 2012; Liu et al.,
2014). Over the past five decades with growing observational
capabilities, it was established that the sunspot structure and
corresponding photospheric magnetic field may change suddenly and
irreversibly after flares. The flare-induced changes are manifested
in the rapid change of the transverse photospheric magnetic field
($B_h$), and darkening as well as decay or even disappearance of
sunspot structures.

In a recent study, Liu et al. (2014) presented a comprehensive
comparison of two major events released from the same NOAA active
region (AR 11283). Both events are characterized by X-class flares
(an X2.1 one on 2011 September 6 and an X1.8 one on September 7),
with fast filament eruptions and coronal mass ejections (CMEs).
The authors found that both flares result in rapid increases of
$B_h$ around the flaring polarity inversion line (PIL) and
decreases in the surrounding peripheral penumbral region,
corresponding to the darkening or decay in white-light (WL)
intensities respectively. This is interpreted as the result of the
inward collapse of the central magnetic field (also see Hudson,
2000) and the radial outward stretching of the peripheral magnetic
field during the flare-CME eruption.

From the space weather forecasting perspective, it is more
important to figure out whether there exist some general trends in
the pre-flare photospheric magnetic field and WL
evolutions of involved sunspots. In Ruan et al. (2014), a study focusing on the 6-hour
long pre-flare sunspot activities and photospheric magnetic field
evolution of the X2.1 event from AR 11283 was presented. It was
concluded that the persistent sunspot rotation plays an important
role in twisting, energizing, and destabilizing the coronal
filament-flux rope system. During the period of apparent sunspot
rotation, it was found that both horizontal field strength ($B_h$) and the inclination angle
($\theta_B$ the angle between the vector magnetic field
and the local radial direction) decline gradually. They found
that the variation of the surface field and the inclination angle
is associated with the overall ascending motion of the corona
filament-flux rope structure.

Ruan et al. (2014) proposed that the long-term pre-flare evolution
of the photospheric $B_h$ can be taken as a possible precursor of
an eruption. In addition, the photospheric field evolution carries
information about the energy storage and triggering process of the
event, and can be used to discern different eruption mechanism. It
was suggested that a gradual decrease of $B_h$ may be a precursor
for an eruption in terms of the flux rope instability (see, e.g.,
Lin et al., 2003), while a persistent increase of this quantity
may be a precursor of the tether-cutting reconnection scenario
(Moore 2001).

In this study, we investigate the X1.8 flare on 2011 September 7,
occurred in the same AR as studied by Ruan et al. (2014). As
mentioned, this event was associated with a fast CME-filament
eruption, very similar to the preceding X2.1 event according to
the Solar Dynamics Observatory (SDO; Pesnell et al., 2012) data.
Our focus is the long-term pre-flare evolution of the photospheric
magnetic field and the sunspot structure, and their correlation
with the pre-eruption dynamics in the upper layers of the solar
atmosphere.

\section{Observations and the overall profile of the event}

For this study, we mainly analyzed the multi-wavelength imaging
data provided by the Atmospheric Imaging Assembly (AIA; Lemen et
al. 2012) and the vector magnetic field and continuum intensity
data by the Helioseismic and Magnetic Imager (HMI; Schou et al.
2012) on board the Solar Dynamic Observatory (SDO). The AIA data
at passbands of 304 \,\AA{} (HeII, T$\sim$0.05 MK), 171 \,\AA{}
(FeIX, $\sim$0.6 MK) and the 94 \,\AA{} (FeXVIII, T$\sim$6.3 MK)
are examined in order to reveal the filament and coronal dynamics
at different temperatures. The processed disambiguated HMI vector
magnetic field data are of 12-minute cadence at a
0.5$^{\prime}$$^{\prime}$ pixel resolution, provided by the HMI
team (see ftp://pail.stanford.edu/pub/HMIvector2/movie/ar1283.mov
for the movie). We also used the HMI continuum intensity
observation at the 6173\,\AA{} to investigate the evolution of the
sunspot structure. These intensity data have been normalized and
removed of limb darkening effect with a cadence of 12 minutes and
a 0.5$^{\prime}$$^{\prime}$ pixel resolution. We also examined the
TiO image (a proxy for continuum at 7057 \,\AA{}) taken by the New
Solar Telescope (NST; Goode et al. 2010; Cao et al. 2010) at Big
Bear Solar Observatory (BBSO) with a high spatial resolution of about
0.04$^{\prime}$$^{\prime}$/pixel.

The AR 11283 was located N14W30 at 19:55 UT on September 7, close
to the disk center. As mentioned, it released two X-class flares
on September 6 and 7, respectively. The eruption processes of
both events have been well-studied by many authors (Feng et al.,
2013; Jiang et al., 2013; Wang et al., 2012; Zharkov et al., 2013;
Ruan et al., 2014; Liu et al. 2014; Shen et al. 2014). So details
of the overall evolution of this AR will not be repeated here.

In Figure 1, we show the GOES SXR (1-8\,\AA{}) light curve from
20:00 UT Sept. 6 to 24:00 UT Sept. 7. The X1.8 flare started at
22:32 UT, peaked at 22:38 UT, and ended at 22:44 UT according to
the GOES data. A C3 flare took place from 19:55 UT to 20:19 UT
with a peaking time at 20:06 UT, which is also of interest to this
study. The peaking times of these three flares have been labeled
with blue vertical dotted lines. According to the CDAW
(Coordinated Data Analysis Workshops) catalog of the LASCO data
(Brueckner et al. 1995), the X1.8 flare was accompanied by a CME
travelling at a linear speed of 792 km s$^{-1}$.

In Figure 2, we present overall structure of this active region at
several observing wavelengths. Panel (a) is the BBSO
high-resolution image taken at 18:15 UT to show details of the
sunspot structure. The image has been rotated by -22.6$^{\circ}$
to facilitate comparison with the SDO data. Panel (b) is
for the HMI 6173 \,\AA{} intensity data in a similar field of view
(FOV) as that of panel (a). Panel (c) is taken from the HMI vector
magnetic field data of this AR with the color map representing the
vertical field component ($B_z$) and arrows for the $B_h$
component. Panels (d) to (f) present the 304, 171, and 94 \,\AA{}
images recorded by SDO around the same time. The contours are
given by $\pm$300~G of the HMI vertical field component ($B_z$) on
the solar surface. This figure shows some pre-eruption condition
of the AR.

The large-scale magnetic field is quadrupolar, with the two spots
on the west giving a $\delta$ configuration, in which the negative
one being the leading and the positive one being the following
spot. Both X-class flares occurred in similar areas in this AR
(Liu et al., 2014). The yellow line in panel (c) delineates an
"L"-shaped PIL. The HMI movie given above provides the long-term
evolution of the sunspot and the magnetic field, from which we
observe that the following $\delta$ spot continues to move
eastward after the X2.1 flare (peaking at 22:20 UT, September 6).
The magnetic field around the PIL is almost parallel to the PIL
indicating a severely-sheared state of the magnetic field. This is
consistent with the transverse alignment of the surrounding
filamentary penumbral structures as seen in panel (a) of Figure 2.

Liu et al. (2014) measured the shearing speed of the two
opposite-polarity spots in this $\delta$ configuration,
and detected a weaker converging motion along the
north-south direction. In addition, after a careful inspection of
the HMI data, we find some signatures of rotation of the positive
$\delta$ spot, although no clear features like a well-developed
magnetic tongue can be used to undoubtedly trace the rotation
(see, e.g., Ruan et al., 2014). No apparent flux emergence is
observed during the period between the X2.1 and the X1.8 flare.
Generally speaking, the above existing sunspot motions
continuously transport energy to the corona through their magnetic
connection, and play a fundamental role in pushing the corona
state to the eruptive point.

The 304-171-94 \,\AA{} images presented in panels (d)-(f) and the
accompanying animation show the pre-eruption structures and
dynamics of the filament and the corona. We can see the filament
is above the PIL with the northern foot rooted in the positive
spot umbra. It has a sigmoidal shape, corresponding to the hot
sigmoidal structure observed with 94 \,\AA{}. The filament-sigmoid
structure has often been taken as the signature of a twisted flux
rope structure (e.g., Rust \& Kumar 1994; Titov \& Demoulin 1999;
McKenzie \& Canfield 2008), which shall carry a major part of the
free magnetic energy to be released.

Now let us inspect the dynamic evolution of the relevant
filament-corona structure. Initially, the filament is buried
underneath the overlying coronal arcades (171\,\AA{}). As the
region evolves, the coronal arcades expand gradually and
continuously, especially those arcades atop the filament. These
arcades seem to be consistently removed from the filament top.
This allows a larger part of the filament to be exposed. Before
the eruption, the filament becomes thicker, darker and more
bulging than before, as seen from both the 304 and 171 \,\AA{}
images. Brightening of loops atop the filament can be seen for
several times from the 171 and 94 \,\AA{} passbands, indicating
reconnections on-going there. The post-flare loops of the C3 flare are
representative of these loops. Later, we will discuss the possible
role of this C3 flare and following reconnections in triggering
the subsequent major eruption.

\section{Correlation of the photospheric transverse field and the penumbral decay}

In Figure 3 we present the long-term temporal evolution of 6173
\,\AA{} intensity map (upper panels) and the photospheric
transverse field ($B_h$, lower panels) of the AR derived from the
HMI data. $B_h$ is illustrated using a color map with red color representing
stronger and blue representing weaker magnetic field strength.

This figure and the accompanying animation exhibit a very striking
feature, that is the close correlation between the gradual changes
of $B_h$ and the WL intensity ($I_c$). This feature is the focus
of our study. Note that for completeness the data pre- and during
the previous X2.1 flare have been included in the animation. Here
we only focus on the data since 00:00 UT of September 7.

Initially, we see that the region with enhanced $B_h$ is allocated
with the PIL. Then, the red color of this region gradually evolves
into a region of yellow, green and blue colors. This indicates
that $B_h$ there becomes weaker with time. After the peaking of
the X1.8 flare, the $B_h$ gets enhanced suddenly. On the other
hand, the sunspot penumbrae around the PIL get brightened
persistently in almost the same period ($>$10 hours before the
X1.8 flare), and become darkened suddenly during the flare. The
changes of $B_h$ and the penumbral structures during the flare are
consistent with previous studies mentioned in our introduction. Of
particular interest here is their long-term pre-flare evolution
and correlation of the two quantities. In the earlier
papers reported the flare-induced penumbra intensity changes
(Wang et al. 2004; Deng et al., 2005), the authors adopted the
wording ``penumbral decay'' to describe the process. Here we
follow them on the terminology, although the wording ``penumbral
fade'' might be a better choice since during the process only the
penumbral intensity changes considerably while the area remains
largely unchanged.

To further examine their evolution and correlation, we select a
trapezoid to include the main area of the pre-flare enhanced $B_h$
region, which is around the PIL across the $\delta$ spot. We then
calculate the averages of the magnetic field components ($B_h$ and
$B_z$), inclination angle ($\theta_B$), total magnetic flux
($\phi$), and the WL intensity ($I_c$) within the trapezoid. The
obtained profiles are plotted in the two panels of Figure 4. In
the upper panel, we see that after $\sim$7:00 UT, $B_h$ starts to
decline gradually till the X1.8 flare, from an initial magnitude
of $\sim$1226 G to a final value of $\sim$679 G. The overall
declining percentage is $\sim$45$\%$. During the X1.8 flare, $B_h$
jumps to $\sim$1053 G. Similarly, $\theta_B$ decreases from
$\sim$68$^{\circ}$ (10:00 UT) to $\sim$55$^{\circ}$ before the
flare, and jumps back to 64$^{\circ}$ after the flare (24:00 UT).
On the other hand, the average intensity starts to increase also
from $\sim$10:00 UT, from a normalized value of $\sim$0.70 to
$\sim$0.89 before the flare and drops to 0.78 after the flare
(24:00 UT). The intimate anti-correlation between $I_c$ and $B_h$
(or $\theta_B$) is self-evident.

In the lower panel of Figure 4, we plot the profiles of the
average positive and negative components of $B_z$ as well as the
total flux ($\Phi$). We see that during the whole pre-flare stage,
all the three quantities do not show any considerable systematic
changes. The value of the positive (negative) $B_z$ lies in a
narrow range of [251, 280] G ( [-174, -215] G), and $\Phi$ lies in
a range of [0.58x$10^{18}$, 0.65x$10^{18}$] Maxwell, from 05:00 UT
to 21:00 UT. This indicates that the PIL region is dominated by
the transverse component of the photospheric field, and there is
no significant flux emergence or cancellation in the period of
study.

A straightforward explanation of the results shown in Figures 3
and 4 is that the magnetic field lines on the photosphere becomes
more vertical with time during the pre-flare phase. To find
further observational support of this explanation, we examine the
AIA data taken in the 304 and 171 \,\AA{} passbands. Following
Ruan et al. (2014), we plot the height-time map along a slice of
the filament. The slice location and the map are shown in the
upper panels of Figure 5, with an accompanying animation available
online. A dashed white line is plotted for visual guide. We see
that, along the slice, at about 08:00 UT the filament becomes
thick enough to be observable from the map. After that, it shows a
gradual ascending motion, very similar to the filament motion
leading to the previous X2.1 flare (Ruan et al., 2014).

The 171 \,\AA{} data present the filament and its overlying
arcades. As described in the previous section, initially the
filament seems to be buried underneath these arcades. Along with
the continuous apparent expansion of the arcades, a larger part of
the filament comes into view. To show this continuous expansion,
we make an angular time map along a circular slice centered around
one foot of the arcades. The northward direction is taken to be 0
degree and the angle increases clockwise. The slice and the map
are shown in the lower panels of Figure 5, where the expansion is
manifested as a gradual rotating motion, which is obvious from
05:00 UT on. The apparent expansion is initially very fast, then
gets slower with time. The outer edge of the loops move from
$\sim$50$^{\circ}$ at 05:00 UT to $\sim$100$^{\circ}$ at 12:00 UT
and reaches $\sim$120$^{\circ}$ from 12:00 UT to 22:00 UT, along
the circular slice. Note that thermal effects, which can make the
coronal arcades visible or invisible, can result in some pseudo
expansion. Nevertheless, with a careful inspection of the
animation, we believe that the genuine physical expansion of the
arcades is important here.

Assuming the pre-eruption corona field evolves slowly enough so that
it can be represented by a series of magnetic equilibria, Liu et
al. (2014) deduced this slow evolution with a Nonlinear Force-Free
Field (NLFFF) extrapolation technique (Wieglemann, 2004;
Wieglemann et al., 2006). In their Figure 4 and the accompanying
animation, they showed the long-term evolution of the extrapolated
electric current density distribution and magnetic field lines
across the central part of the filament. It is clear from their
study that there exists a flux rope aligning with the filament and
the flux rope arises gradually during the period from after the
X2.1 to before the X1.8 flare. This result is consistent with the
observation that a significant part of the well-developed filament
exists after the X2.1 flare (see also Liu et al. 2014), and
supports our argument deduced above.

In summary, we found a persistent well-correlated pre-flare long
term evolution of the penumbral structure and the photospheric
transverse field around the PIL. The observed penumbral decay and
$B_h$ decline are likely caused by the ascending motion of the
filament-flux rope system. This picture is supported by the
simultaneous imaging data with the 304 \,\AA{} for the filament
and the 171 \,\AA{} for the overlying arcades. Nevertheless, it
cannot be ruled out of the possibility that the pre-flare sunspot
decay and the $B_h$ decline are simply the result of the overall
sunspot evolution and not the direct response to the
filament-corona dynamics.

\section{Possible triggering process of the major eruption}

Another interesting problem to discuss here is the possible role
played by the C3 flare in triggering the following X1.8 eruption.
To show this, in Figure 6 we present the four AIA 171 \,\AA{}
images from 20:39 UT to 22:27 UT. The first panel presents the
post loops of the C3 flare, stretching over the filament. A nearby
bright loop on the northern side is also shown. In the second
panel both loops get darkened and expanded. In the third panel
(21:54 UT) a new loop connecting the southern foot of the
post-flare loops and the eastern foot of the nearby loop appears.
This is very likely a result of the reconnection between the two
loop systems, as will be explained in the following paragraph. At
22:27 UT, in the fourth panel, the post-reconnection loops get
brightened considerably and the post-C3 loops almost disappear.
This evolution of coronal loops can also be clearly seen
from the accompanying AIA 171 \,\AA{} animation.

To support this reconnection scenario, we show the contours of the
vertical magnetic field component of HMI in Figure 6 (b). It is
seen that the magnetic field polarities are in the order of
positive, negative, positive, and negative for the four foots of
the two sets of pre-reconnection loops, thus favoring the
reconnection process described above. The
post-reconnection bright loops (see Figure 6d) are still likely on
top of the filament and more high-lying in the corona in
comparison to those pre-reconnection overlying loops. Shortly
after the reconnection, the filament starts to rise rapidly and
the X1.8 flare takes place.

To illustrate the reconnection process more clearly, we show
schematics in the left two panels at the bottom of Figure 6, where
yellow lines are for pre-reconnection loops and red lines are for
post-reconnection loops. It can be seen that the expansion of the
coronal arcades overlying the filament (in cyan) drives their
reconnection with longer magnetic loops rooted in a nearby pair of
opposite polarities. This reconnection can remove some of the
arcades overlying the filament and reduce the confining force (or
the strapping effect, see recent paper of Wang et al., 2015)
acting on the filament. This will at least speed up the evolution
process leading to the following major eruption, and possibly play
as a trigger of the filament-flux rope instability.
\textbf{Possible configuration during the impulsive stage of the
X1.8 flare is also shown in the last panel of Figure 6, for
completeness, from which we see that the filament erupts with a
flux rope and part of its overlying arcades.} The picture involves
a quadrupolar topology and a triggering reconnection in the
corona, both are essential features of the breakout model
(Antiochos et al., 1999) for solar eruption. We therefore suggest
that the breakout process may be important to the trigger of the
X1.8 flare and the associated CME.

\section{Conclusions and discussion}

Some recent studies on the magnetic field evolution have been
focusing on rapid changes of the photospheric transverse field and
sunspot structures induced by solar flares. In this study we
investigate the correlation between the long-term ($\sim$10-20
hours) pre-flare evolution of the sunspot penumbrae and the
photospheric transverse field component around the PIL, as well as
their relation with filament-corona dynamics. It is found that the
penumbrae decayed gradually and the strength of the transverse
field (and the inclination angle of the magnetic field) on the
solar surface declined correspondingly, indicating that the
pre-flare magnetic structure from the photosphere to the corona is
getting more vertical with time. This indication is corroborated
with the SDO imaging observation of the filament arising and the
apparent expansion of its overlying arcades, and consistent with
the NLFFF extrapolation results of the pre-eruption state of the
local corona.

Intensity changes of the sunspot penumbra induced by solar
flares have been reported in several studies (e.g., Wang et al.,
2004; Deng et al., 2005; Liu et al., 2014). Here we report the
long-term pre-flare evolution of the penumbra intensity change as
well as its correlation with the photospheric magnetic field
evolution and relevant coronal dynamics. In Deng et al. (2005),
the flare-induced penumbra decay was explained with the change of
the inclination angle of the magnetic field in the corresponding
penumbral region. According to Leka {\&} Skumanich (1998), the
magnetic field inclination angle in the peripheral penumbrae, when
turning from more inclined to more vertical and toward the umbra,
can directly suppress the penumbral Evershed flow resulting in an
increase of the continuum intensity. Here we propose a very
similar scenario to interpret the observational result, although the trend
of long term change is opposite to that of the flare-induced rapid changes.
We suggest that the observed persistent pre-flare penumbral decay (or fade)
is a result of the gradual change of the direction of the magnetic
field in the penumbral region from more horizontal to more
vertical, which is likely caused by the gradual rising of the
filament-flux rope system in the upper solar atmosphere.

The long-term pre-flare behavior of the photospheric magnetic
field, sunspot, filament and the corona arcades are of vital
importance to space weather studies. These behaviors could provide
clues about how the solar magnetic field evolves from a
pre-eruption state to the eruption, including the energy
transport, storage, and release processes. The result of our study
is based on eruptions from only one AR. More events
should be investigated to decide what signatures or processes
could be used as possible eruption precursor or trigger.

\begin{figure*}[!htbp]
\vspace{12.3mm}
\centering
\includegraphics[width=120mm]{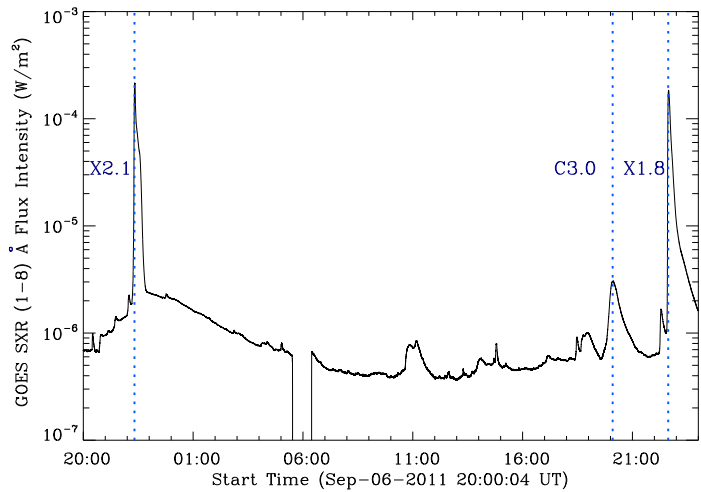}
\caption{The 1-8 \,\AA{} GOES SXR flux intensity profiles from
20:00 UT September 6 to 24:00 UT September 7. The peaking times of
these three flares X2.1 (22:20 UT), C3.0 (20:06 UT), X1.8 (22:38
UT) have been labelled with blue vertical dotted lines. An
animation and a color version of this figure are available online.}
\label{fig:bright}
\end{figure*}

\begin{figure*}[!htbp]
\centering
\includegraphics[width=140mm]{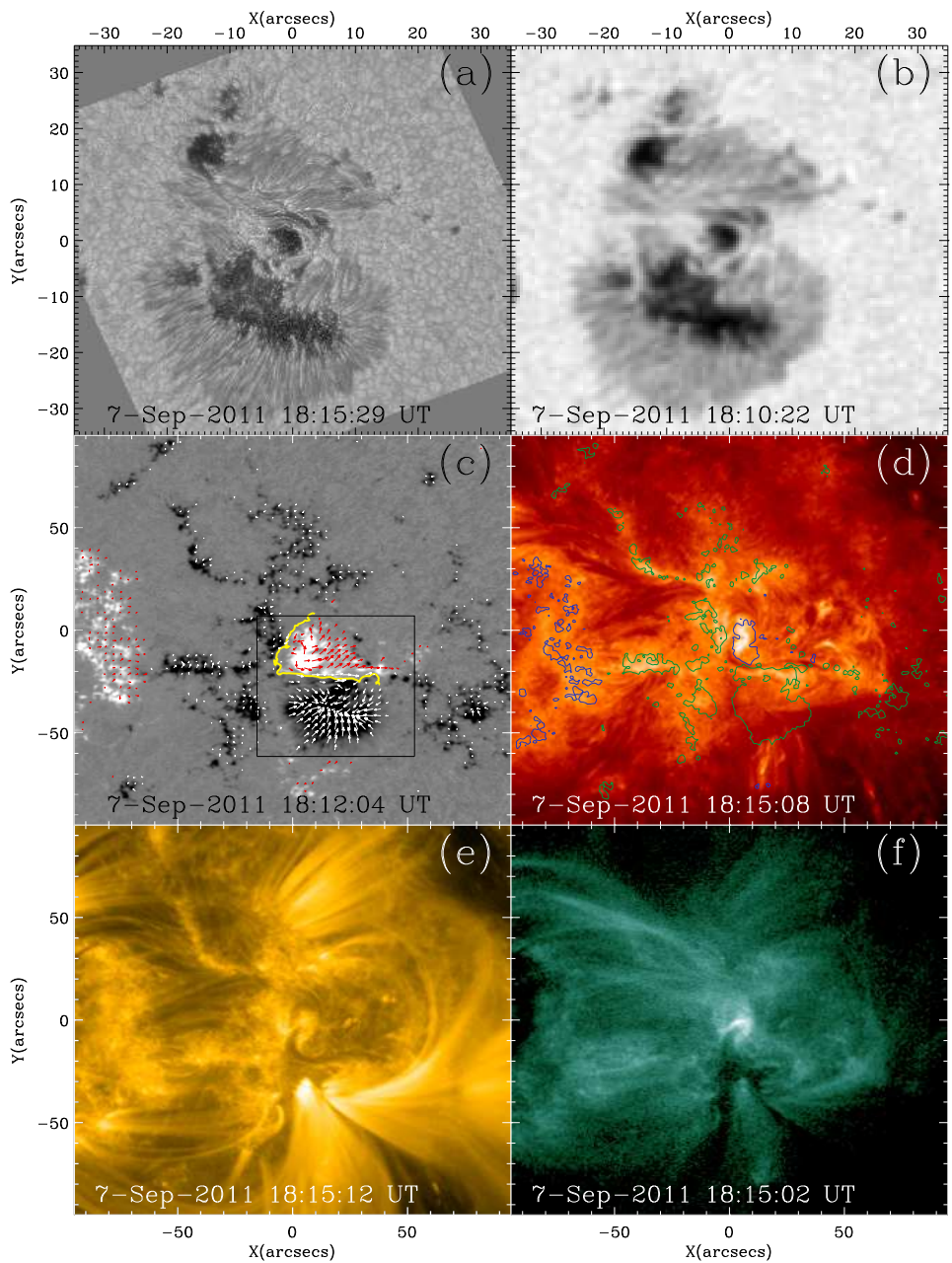}
\caption{Pre-eruption observations of the AR 11283: (a) the BBSO
high-resolution image taken at 18:15 UT, (b) the HMI 6173 \,\AA{}
intensity image, (c) the HMI vector magnetic field with the color
map representing the vertical field component ($B_z$) and arrows
for the $B_h$ component, (d) to (f) the 304, 171, and 94 \,\AA{}
images recorded by SDO. The contours in panel (d) are given by
$\pm$300 G of the HMI vertical field component ($B_z$) on the
solar surface. The FOV of panels (a) and (b) has been marked in panel (c) with a square.
A color version of this figure is
available online.} \label{fig:bright}
\end{figure*}

%%%Fig 3
\begin{figure*}[!htbp]
\centering
\includegraphics[width=140mm]{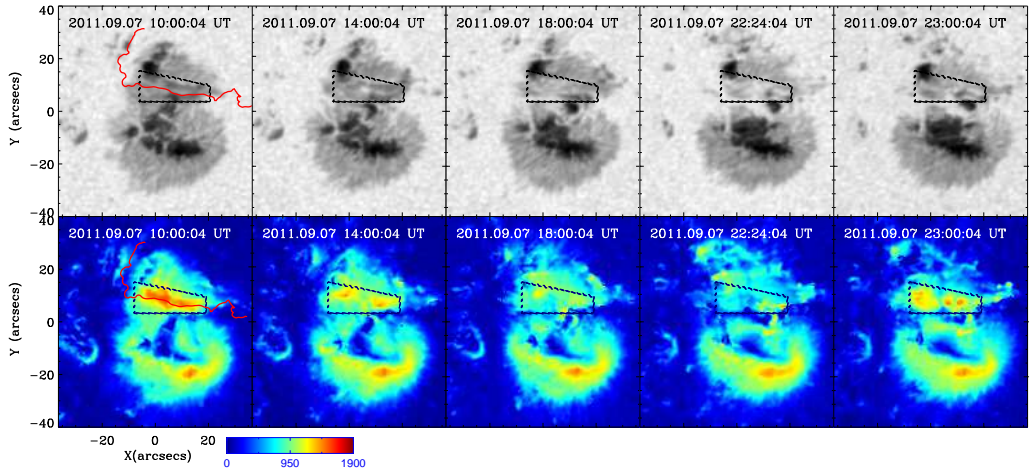}
\vspace{12.3mm}
%%%\end{minipage}
%\hspace*{\fill}
\caption{Temporal evolution of 6173 \,\AA{} intensity map (upper
panels) and the photospheric transverse field strength ($B_h$, lower
panels) of the AR given by HMI. An animation and a color version
of this figure are available online.} \label{}
\end{figure*}

\begin{figure*}[!htbp]
%\hspace*{12.3mm
%\begin{minipage}{\textwidth}
%\includegraphics[width=0.49\textwidth,angle=0,clip]{f4a.ps}
%\hspace{-6.0mm}
%\includegraphics[width=0.49\textwidth,angle=0,clip]{f4b.ps}
%\end{minipage}
\includegraphics[width=140mm]{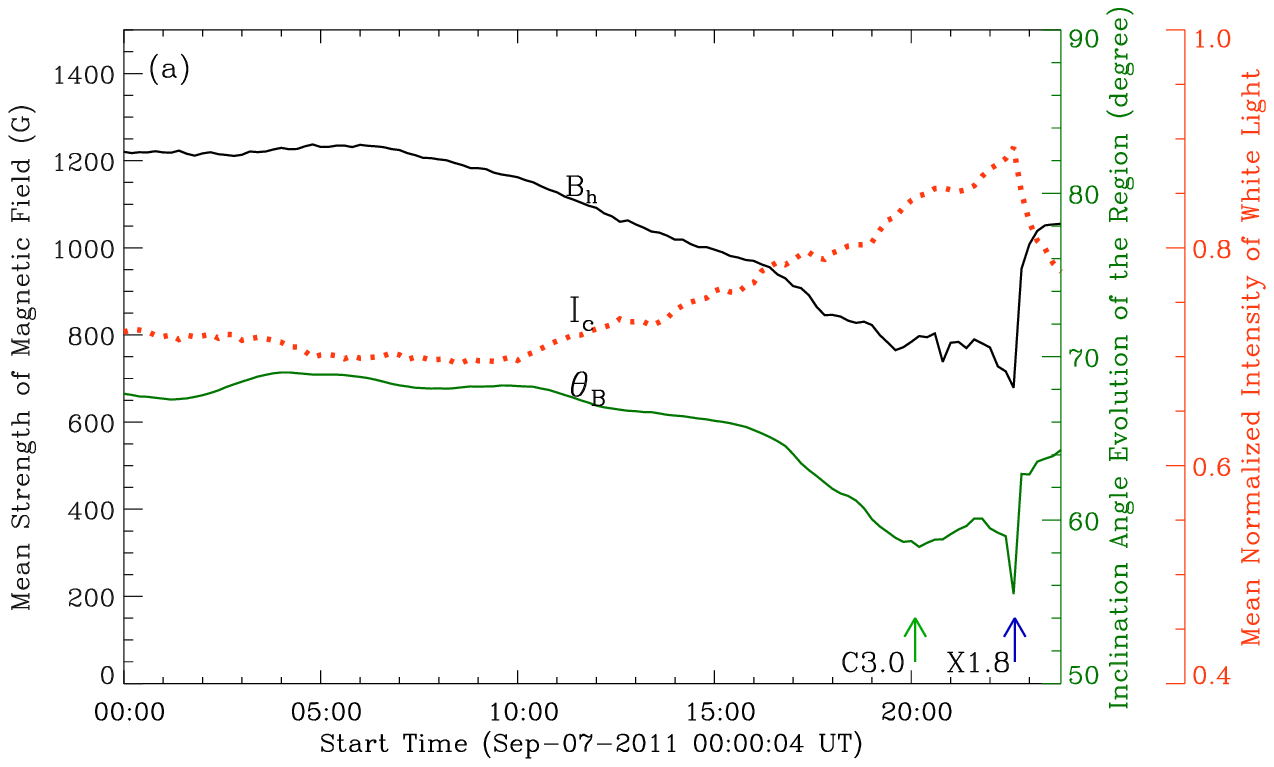}
\includegraphics[width=140mm]{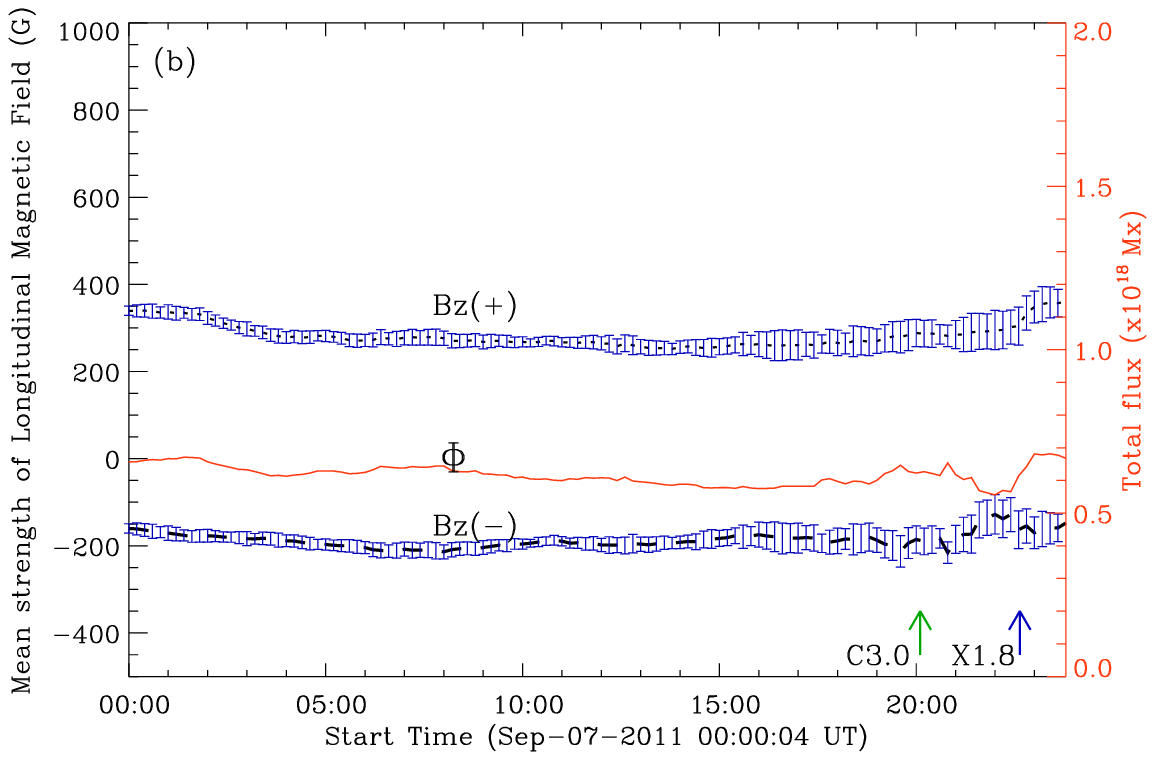}
\caption{(a) Temporal profiles of the average transverse field strength
$B_h$, average inclination angle $\theta_B$ and normalized
white-light intensity $I_c$, and (b) Temporal profiles of the
average positive and negative components of $B_z$ as well as the
total flux ($\Phi$). All parameters are calculated in the area
defined by the trapezoid shown in Figure 3. A color version of
this figure is available online.} \label{}
\end{figure*}

\begin{figure*}[!htbp]
\centering
\includegraphics[width=140mm]{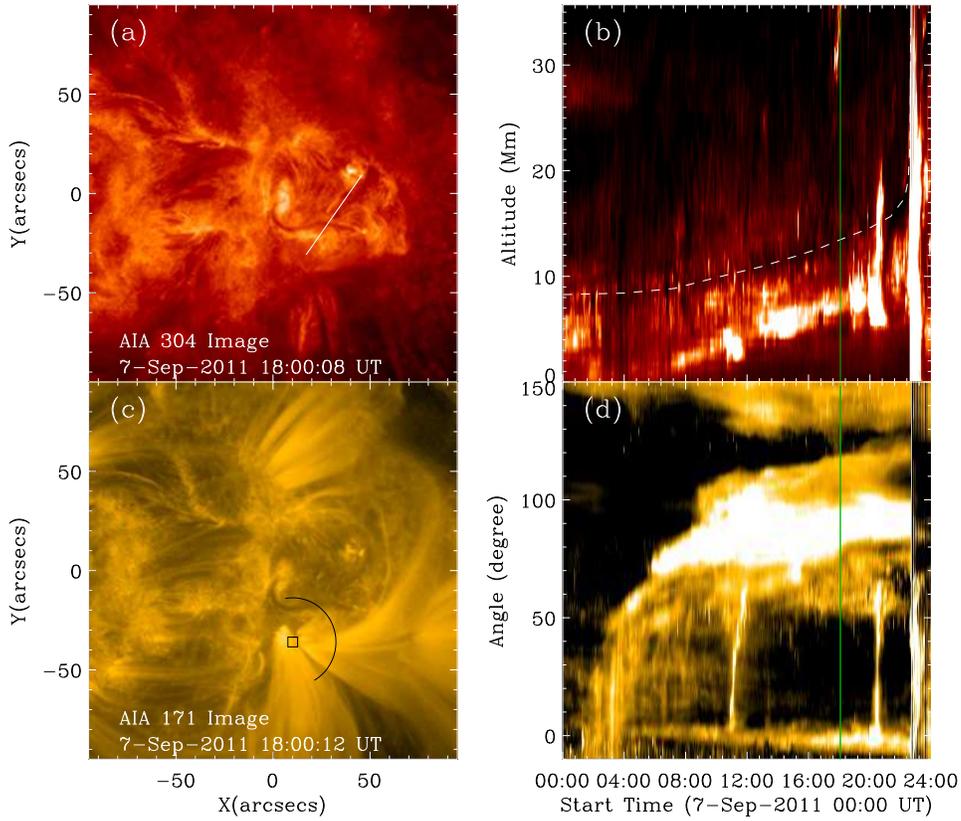}
\caption{Time-stacked images along specified slices of the
filament and the overlying arcades. Panels (a) and (c) show the
direct images of AIA/SDO at 18:00 UT on September 7, 2011 in the
304\,\AA{} and 171\,\AA{} passbands, with the location of the
slices. The green vertical lines in (b) and (d) represent the
observation times of panels (a) and (c). An animation and a color
version of this figure are available online.} \label{fig:bright}
\end{figure*}

\begin{figure*}[!htbp]
\begin{minipage}{\textwidth}
\includegraphics[width=140mm]{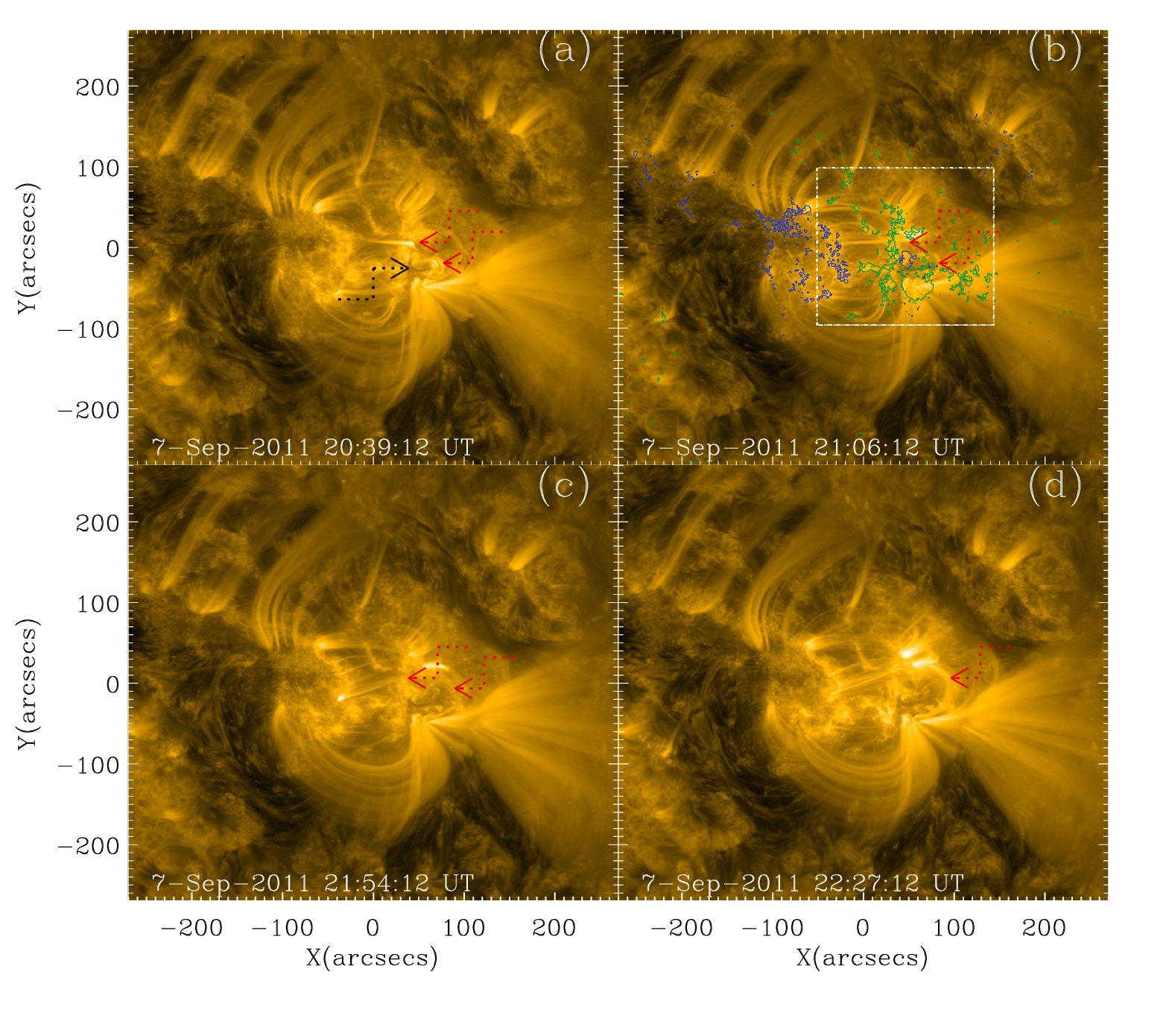}
\end{minipage}
\vspace{-7.0mm}

\begin{minipage}{\textwidth}
\hspace{2.7mm}
%\vspace{-10.0mm}
\includegraphics[width=130mm]{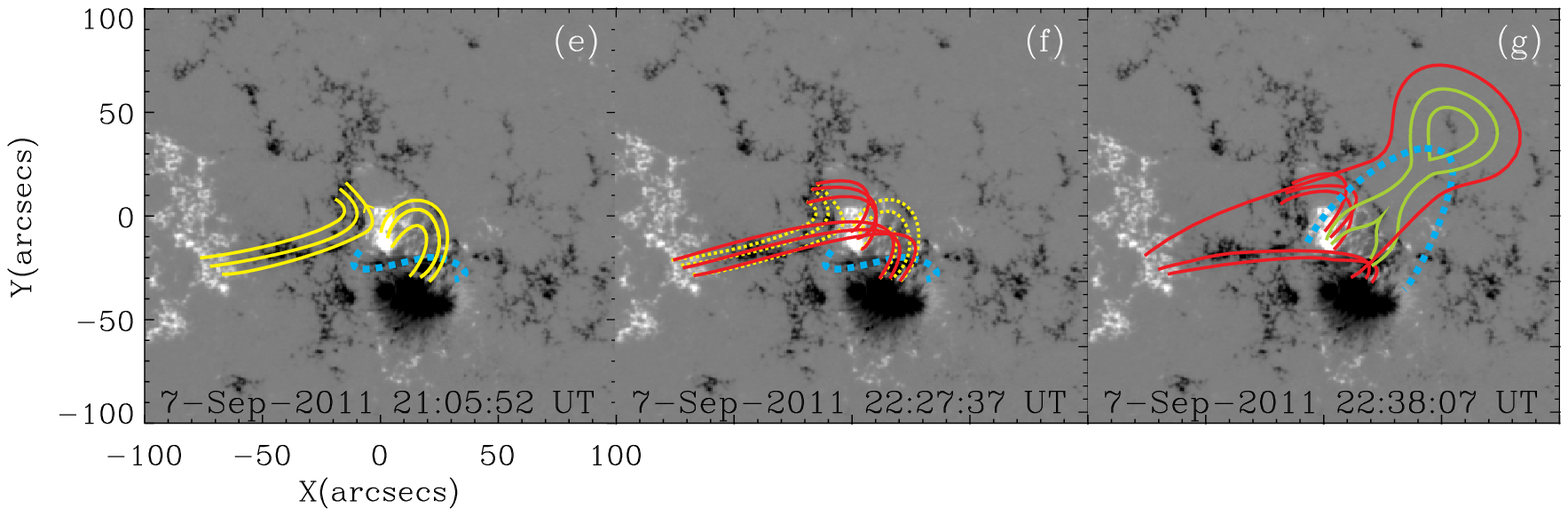}
\end{minipage}
\caption{Panels (a-d): The AIA 171 \,\AA{} images from 20:39 UT to
22:27 UT. The contours in panel (b) are given by 300 G for the
positive component (in blue) and 200 G for the negative component
(in green) of $B_z$ on the solar surface. Red arrows point to the
loops of interest, and the black arrow in panel (a) points to the
filament of study. Panels (e-g): Schematics showing the proposed
reconnection-triggering process \textbf{and possible magnetic
configuration during the impulsive stage of the major eruption.}
The pre-reconnection magnetic field lines are in yellow and
post-reconnection lines are in red, and the filament of study is
depicted with a cyan dashed line. Arrows in (e) indicate the
magnetic field direction. An animation and a color version of this
figure are available online.} \label{fig:bright}
\end{figure*}

\newpage
\acknowledgements We thank SDO/HMI and SDO/AIA science teams for
the free access to the data. We are grateful to the referee for valuable comments
and Dr. Guohui Du, Bing Wang, and Di Zhao for their help in preparing the figures.
This work was supported by the 973 program NSBRSF 2012CB825601,
U1331104 and NNSFC grants 41331068, 41274175 to SDUWH, and by NSF
grants AGS-1348513 and AGS-1408703 to NJIT.

\end{document}